\documentclass[runningheads]{llncs}
\usepackage{graphicx}
\usepackage{color}
\usepackage{hyperref}

\usepackage{makecell}
\usepackage{adjustbox}
\usepackage{threeparttable}
\usepackage{multirow}
\usepackage{amsmath}
\usepackage{bm}
\usepackage{lipsum}
\usepackage[misc]{ifsym}

\begin{document}
\title{Training Automatic View Planner for Cardiac MR Imaging via Self-Supervision by Spatial Relationship between Views}
%
\titlerunning{Automatic View Planning for Cardiac MR Imaging via Self-Supervision}
%
%
\author{Dong Wei\textsuperscript{(\Letter)} \and Kai Ma \and Yefeng Zheng}
\authorrunning{D. Wei et al.}
%
\institute{Tencent Jarvis Lab, Shenzhen, China\\
\email{\{donwei,kylekma,yefengzheng\}@tencent.com}}
\maketitle              
\begin{abstract}
View planning for the acquisition of cardiac magnetic resonance imaging (CMR) requires acquaintance with the cardiac anatomy and remains a challenging task in clinical practice.
Existing approaches to its automation relied either on an additional volumetric image not typically acquired in clinic routine, or on laborious manual annotations of cardiac structural landmarks.
This work presents a clinic-compatible and annotation-free system for automatic CMR view planning.
The system mines the spatial relationship---more specifically, locates and exploits the intersecting lines---between the source and target views, and trains deep networks to regress heatmaps defined by these intersecting lines.
As the spatial relationship is self-contained in properly stored data, \emph{e.g.}, in the DICOM format, the need for manual annotation is eliminated.
Then, a multi-view planning strategy is proposed to aggregate information from the predicted heatmaps for all the source views of a target view, for a globally optimal prescription.
The multi-view aggregation mimics the similar strategy practiced by skilled human prescribers.
Experimental results on 181 clinical CMR exams show that our system achieves superior accuracy to existing approaches including conventional atlas-based and newer deep learning based ones, in prescribing four standard CMR views.
The mean angle difference and point-to-plane distance evaluated against the ground truth planes are 5.98$^\circ$ and 3.48 mm, respectively.

\end{abstract}
\section{Introduction}
Cardiac magnetic resonance imaging (CMR) is the gold standard for the quantification of volumetry, function, and blood flow of the heart \cite{la2012cardiac}.
A great deal of effort is put into the development of algorithms for accurate, robust, and automated analysis of CMR images, with a central focus on the segmentation of cardiac structures \cite{bai2019self,painchaud2019cardiac,robinson2017automatic,wei2015medical,wei2013comprehensive,zotti2018convolutional}.
However, much less attention has been paid to automatic view planning for the acquisition of CMR, which still remains challenging in clinical practice.
First, imaging planes of CMR are customized for each individual based on specific cardiac structural landmarks \cite{kramer2020standardized},
and the planning process demands specialist expertise \cite{suinesiaputra2015quantification}.
Second, the planning process adopts a multi-step approach involving several localizers defined by the cardiac anatomy, which is complex, time-consuming, and subject to operator-induced variations \cite{lu2011automatic}.
These factors may constrain the use of CMR in clinical practice. 
Hence, automatic planing system is expected to increase the impact of the specific imaging technology on care of patients suffering from cardiovascular diseases.

A few works attempted automatic view planning for CMR \cite{alansary2018automatic,blansit2019deep,frick2011fully,lu2011automatic}.
For example, both Lu \emph{et al.} \cite{lu2011automatic} and Frick \emph{et al.} \cite{frick2011fully} tackled this challenging task from the perspective of classical atlas-based methods by fitting triangular mesh-based heart models into a 3D volume.
Later, Alansary \emph{et al.} \cite{alansary2018automatic} proposed to employ reinforcement learning to prescribe the standard four-chamber long-axis CMR view from a thoracic volume.
A common foundation of these works was the use of a 3D MRI volume from which the standard CMR views were prescribed.
However, such a 3D volume is not typically acquired in current clinic routine,
where the standard CMR views are sequentially optimized based on a set of 2D localizers.
In an effort to develop a clinic-compatible system, Blansit \emph{et al.} \cite{blansit2019deep} proposed to sequentially prescribe standard CMR views given a vertical long-axis localizer (also known as the pseudo two-chamber (p2C) localizer), driven by deep learning based localization of key landmarks.
However,
this method relied on extensive manual annotations to train the deep convolutional neural networks (CNNs).
Besides,
the p2C localizer---which was the starting point of the entire system---was assumed given.
Yet in practice it requires expertise in cardiac anatomy to prescribe this localizer from scout images in normal body planes (such as the axial view), which is an obstacle to a fully automatic workflow.

In this work, we propose a clinic-compatible automatic view planning system for CMR.
Similar to \cite{blansit2019deep}, our system takes advantage of the power of deep CNNs, but eliminates the need for annotation via self-supervised learning \cite{bai2019self,jing2020self,zhou2019models}.
Above all, we make a critical observation that the way how the existing CMR data have been prescribed is self-contained in correctly recorded data, \emph{e.g.}, in the Digital Imaging and Communications in Medicine (DICOM) format, in the form of the spatial relationship between views.
Then, inspired by the recent progress in keypoint-based object detection \cite{duan2019centernet,law2018cornernet,zhou2019objects,zhou2019bottom}, we propose to regress the intersecting lines between the views, which can be readily computed using the spatial relationship.
Training the networks to predict these intersecting lines actually teaches them to reason about the key cardiac landmarks that defined these lines when the operators produced the CMR data, while at the same time eliminates the need for manual annotation.
After the intersecting lines are predicted in the localizers, the standard CMR views are eventually planned by aggregating the predictions in multiple localizers to search for a globally optimal prescription.

In summary, our contributions are three folds.
First, we propose a CNN-based, clinical compatible system for automatic CMR view planning, which eliminates the need for manual annotation via self-supervised learning of the spatial relationship between views.
Second, we propose to aggregate multi-view information for globally optimal plane prescription, to mimic the clinical practice of multi-view planning.
Third, we conduct extensive experiments to study the proposed system and demonstrate its competence/superiorty to existing methods.
In addition, we demonstrate prescription of the cardiac anatomy defined localizers (including the p2C) given the axial localizer, bridging the gap between normal body planes and CMR views for a more automated workflow.

\section{Methods}
\subsubsection{Preliminary}
Different from the commonly used axial, sagittal, or coronal plane oriented with respect to the long axis of the body (the body planes), CMR adopts a set of double-oblique, cardiac anatomy defined views customized for each individual.
Often these views are prescribed along the long axis (LAX) or short axis (SAX) of the left ventricle (LV) and with respect to specific cardiac structural landmarks
(\emph{e.g.}, the apex and mitral valve),
for optimal visualization of structures of interest and evaluation of cardiac functions.
The mostly used standard CMR views include the LV two-chamber (2C) LAX, three-chamber (3C) LAX, four-chamber (4C) LAX, and SAX views, which provide complementary information for comprehensive evaluation of the heart.
In clinical practice, the 2C and 4C LAX views are also called the vertical long-axis (VLA) and horizontal long-axis (HLA) views, whereas the 3C LAX view is called the left ventricular outflow tract (LVOT) view.

Our imaging protocol generally follows \cite{kramer2020standardized} for LV structure and function.
After adjusting the heart to the isocenter of the bore, a multi-slice axial dark-blood localizer is firstly prescribed through the chest from sagittal and coronal scouts, serving as the basis for the following cardiac anatomy oriented views.
Next, a single-slice pseudo 2C (p2C) localizer---the first cardiac anatomy oriented view---is prescribed from the axial localizer.
Then, a single-slice pseudo 4C localizer (p4C) and a multi-slice pseudo SAX (pSA) localizer are prescribed based on the p2C localizer.
Although similarly defined by the cardiac anatomy, these localizers usually cannot provide accurate anatomic and functional characterization of the heart like the standard views due to the obliquity of the heart walls to the body planes \cite{ginat2011cardiac}, hence are often referred to as the ``pseudo'' LAX (pLA) and SAX views \cite{10.1093/eurheartj/ehw680}.
Eventually, the p2C, p4C, and pSA localizers are used to sequentially optimize the imaging planes for the standard 2C, 3C, 4C, and SAX views.
As in the literature \cite{alansary2018automatic,blansit2019deep,frick2011fully,lu2011automatic}, we aim at planning the standard LAX and SAX views (only the most basal SAX plane is considered) in this work.
In addition, we investigate the prescription of the pseudo-view localizers from the axial localizer.
As the first step of deriving cardiac anatomy oriented imaging planes from the body planes, it
is indispensable for a clinic-compatible system yet
remains uninvestigated in existing literature \cite{blansit2019deep}.

For clarity, we refer to a view plane to prescribe as the \emph{target} plane, and the view(s) leading to the specific target plane as its \emph{source} view(s).
The source to target mappings for the standard LAX and SAX views are shown in Fig. \ref{fig:framework} (see supplement Fig. \ref*{figS:loc} and Table \ref*{tabS:view_dependency} for more details).

\begin{figure}[!t]
\centering
\includegraphics[width=\textwidth,,trim=0 3 0 0,clip]{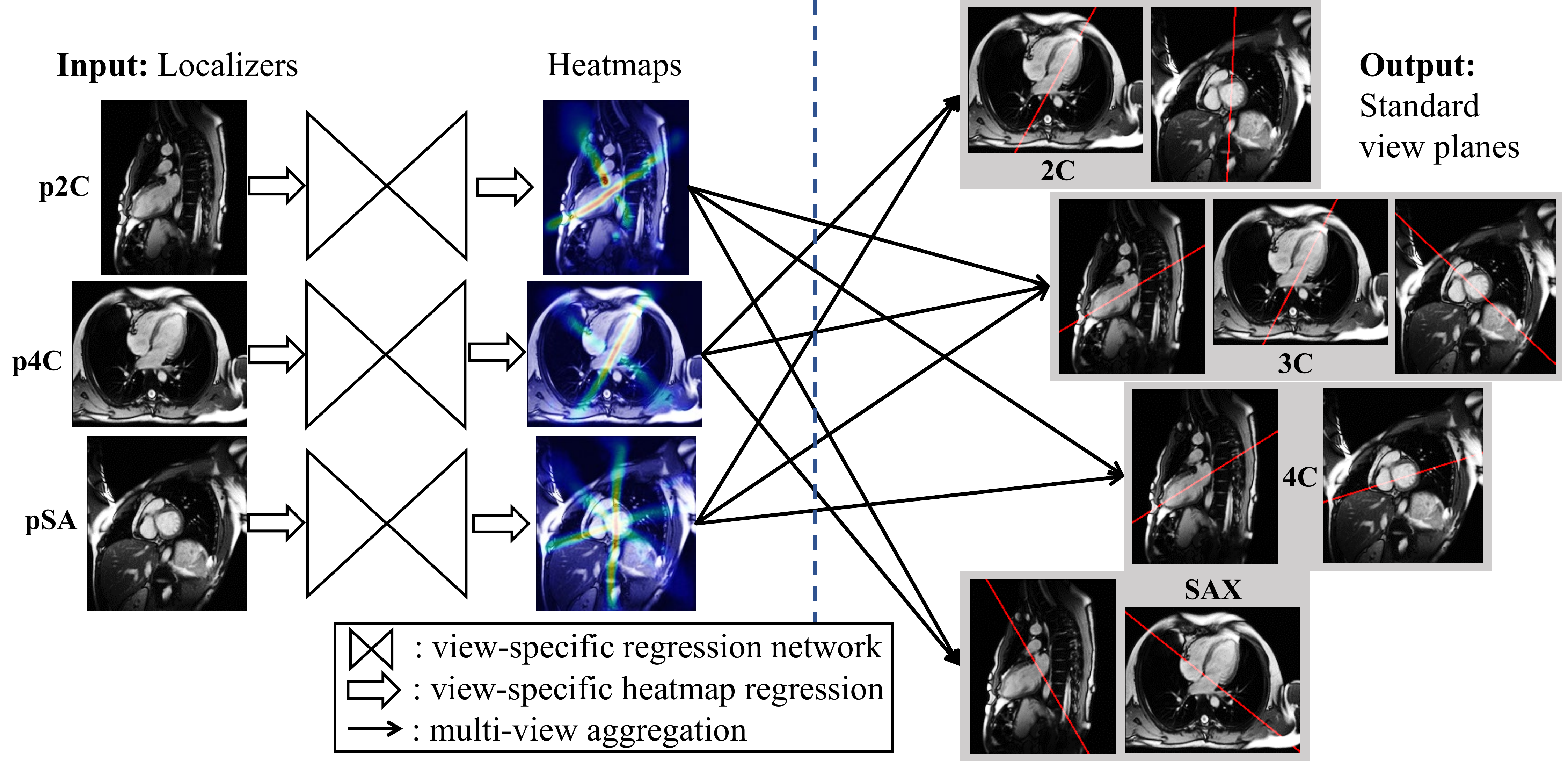}
\caption{Overview of the proposed approach to automatic CMR view planning (illustrated with the task of planning the standard CMR view planes from the localizers).
Left: prediction of standard view planes in localizers via heatmap regression.
Right: prescription of standard view planes by aggregating heatmaps predicted in multiple localizers; the prescriptions are presented as intersecting lines (red) with the localizers.} \label{fig:framework}
\end{figure}

\subsubsection{Method Overview}
The overview of our proposed system is shown in Fig. \ref{fig:framework}, including two main steps.
First, given a set of localizers as input, heatmaps for the target planes are predicted by the view-specific regression networks.
Second, multiple heatmaps for a specific target plane are aggregated to prescribe the target plane as output.
Below we elaborate the two steps.

\subsubsection{Target Plane Regression with Self-Supervised Learning}
A previous work on CNN-based automatic CMR view planning regressed locations of the cardiac structural landmarks, then prescribed the view planes by connecting the landmarks \cite{blansit2019deep}.
A drawback was that the landmark regression networks needed extensive annotations for training.
In this work, we propose to mine the spatial relationship among the CMR data, and directly regress the intersecting lines of the target plane with the source views.
As the spatial information of each CMR slice is recorded in its data header (\emph{e.g.}, of the DICOM format), we can readily compute the intersecting lines.
Therefore, the supervising ground truth of our proposed regression task is self-contained in the data, requiring no manual annotation at all.
In addition, the intersecting lines are actually the prescription lines defined by the operator using cardiac landmarks during imaging.
Hence, our networks are still trained to learn the cardiac anatomy by the proposed regression task.

For practical implementation, we train the networks to regress a heatmap
defined by the distance to the intersecting line, instead of the line itself.
This strategy is commonly adopted in the keypoint detection literature
\cite{duan2019centernet,law2018cornernet,pfister2015flowing,zhou2019objects,zhou2019bottom}, where the regression target is the heatmaps defined by the keypoints.
The benefits of regressing heatmaps include better interpretability and ease of learning for the network \cite{pfister2015flowing}.
Formally, denoting the equation of the intersecting line in the \emph{2D image coordinate system of a source view} by $Ax+By+C=0$, where $(x,y)$ are coordinates and $(A, B, C)$ are parameters, the heatmap is computed as
\begin{equation}\label{eq:heatmap}
    H(x,y)=\exp\big[-{(Ax+By+C)^2}\big/\big(2\sigma^2(A^2+B^2)\big)\big],
\end{equation}
where $\sigma$ is a hyperparameter denoting the Gaussian kernel size.
We define $\sigma$ with respect to the slice thickness of the target view for concrete physical meaning, and study its impact later with experiments.
An L2 loss is employed to train the network:
\begin{equation}\label{eq:L_heat}
    \mathcal{L} = \frac{1}{T}\frac{1}{|\Omega|}
    {\sum}_{t=1}^T{\sum}_{(x,y)\in\Omega}
    \big\|H_t(x,y) - \hat{H}_t(x,y)\big\|^2,
\end{equation}
where $T$ is the total number of target planes that are prescribed from a source view (also the number of channels of the network's output), $(x,y)$ iterates over all pixels in the source view image domain $\Omega$, and $\hat{H}$ is the heatmap predicted by the network (see Fig. \ref{fig:framework} for example predictions by trained networks).


\subsubsection{Plane Prescription with Multi-View Aggregation}
After the heatmaps are predicted for all the source views of a specific target plane, we prescribe the target plane by aggregating information from all these heatmaps.
Specifically, by prescribing the plane with the greatest collective response aggregated over its intersecting lines with all the source heatmaps, we mimic the clinical practice where skilled operators plan the CMR views by checking the locations of a candidate plane within multiple source views.
Let us denote a candidate plane by $P:(p, \theta, \phi)$, where $p$ is a point on the plane, $\theta$ and $\phi$ are the polar and azimuthal angles of its normal in the spherical coordinate system;
and denote the intersecting line segment between $P$ and the $v$\textsuperscript{th} source view (of $V$ total source views) by the exhaustive set of pixels lying on the line within the source view image $I_v$: $l_v=\{(x, y)|(x, y)\in P\cap I_v\}$, where $P\cap I_v$ denotes the intersection.
Then, the optimal target plane can be formally expressed as
\begin{equation}\label{eq:grid_search}
 \hat{P}=\operatorname{argmax}_{(p,\theta,\phi)}
 {\sum}_{v=1}^{V}{\sum}_{(x,y)\in l_v}\hat{H}_v(x,y).
\end{equation}
We propose to grid search for the optimal triplet $(\hat{p},\hat{\theta},\hat{\phi})$ with two strategies to reduce the search space.
First, we constrain the search for $\hat{p}$ along a line segment with which the target plane should intersect.
In practice, the intersecting line segment of a source view with another source view is used.
Second, a three-level coarse-to-fine pyramidal search is employed, with the steps set to 15, 5, and 1 pixel(s) (for $p$) and degree(s) (for $\theta$ and $\phi$).
Note that our multi-view aggregation scheme naturally takes into account the networks' confidence in the predictions, where higher values in the regressed heatmaps indicate higher confidence.


\section{Experiments}

\subsubsection{Dataset and Evaluation Metrics}
With institutional approval, we retrospectively collected 181 CMR exams from 99 infarction patients.
Of these patients, 82 had two exams and the rest had only one.
The exams were performed on a 1.5T MRI system;
details about the image acquisition parameters are provided in supplement Table \ref*{tabS:protocol}.
The dataset is randomly split into training, validation, and test sets in the ratio of 64:16:20;
both exams of the same patient are in the same set.
The validation set is used for empirical model optimization such as hyperparameter tuning.
Then, the training and validation sets are mingled to train a final model with optimized settings for evaluation on the held-out test set.
Following existing literature \cite{alansary2018automatic,blansit2019deep,frick2011fully,lu2011automatic}, we use the angular deviation and point-to-plane distance as evaluation metrics.
Specifically, the absolute angular difference between the normals to the automatic and ground truth planes is computed, and the distance from the center of the ground truth view (image) to the automatic plane is measured.
For both metrics, smaller is better.

\subsubsection{Implementation}
The PyTorch \cite{steiner2019pytorch} framework (1.4.0) is used for all experiments.
We use the U-Net \cite{ronneberger2015u} as our backbone network.
A model is trained for each of the four localizers (axial, p2C, p4C, and pSA).
For stacked localizers (the axial and pSA), all slices of a patient are treated as a mini batch with the original image size as input size, thus both the mini batch size and input size vary with individual.
For the others, we use a mini batch size of eight images, whose sizes are unified according to data statistics by cropping or padding, where appropriate, for training;
specifically, the p2C and p4C localizers are unified to 192$\times$176 (rows by columns) pixels and 160$\times$192 pixels, respectively.
The Adam~\cite{kingma2014adam} optimizer is used with a weight decay of 0.0005.
The learning rate is initialized to 0.001 and halved when the validation loss does not decrease for 10 consecutive epochs.
The exact numbers of training epochs vary with the specific localizer and value of $\sigma$, and are determined with the validation set (range 75--250).
Online data augmentation including random scaling ($[0.9, 1.1]$), rotation ($[-10^\circ, 10^\circ]$), cropping, and flipping (left-right for the p2C, upside-down for the p4C, and none for stacked localizers) is conducted during training to alleviate overfitting.
For simple preprocessing, the z-score standardization (mean subtraction followed by division by standard deviation) is performed per localizer per exam.
A Tesla P40 GPU is used for model training and testing.
The source code is available at: \url{https://github.com/wd111624/CMR_plan}.

\begin{table}[!t]
\caption{Impact of multi-view planning on prescription accuracy on the validation set (with $\sigma=0.5t$).
`-' indicates that a specific plane cannot be prescribed solely from the corresponding subset of localizers (loc.).
Data format: mean $\pm$ standard deviation.}\label{tab:multiview}
\centering
\setlength{\tabcolsep}{1.55mm}
\begin{adjustbox}{width=.95\textwidth}
\begin{tabular}{ccccccccc}
    \hline\hline
    \multicolumn{1}{c}{Target} & & \multicolumn{3}{c}{Normal deviation ($^{\circ}$)} & & \multicolumn{3}{c}{Point-to-plane distance (mm)} \\
    \cline{3-5}\cline{7-9}
    \multicolumn{1}{c}{plane} & Loc.$=$ & pLA & pSA & pLA \& pSA & & pLA & pSA & pLA \& pSA \\
    \hline
    {2C} & & - & 4.96$\pm$2.77 & 4.86$\pm$2.95 & & - & 3.33$\pm$2.33 & 2.78$\pm$1.92 \\
    {3C} & & 41.88$\pm$29.98 & 7.68$\pm$4.67 & 7.18$\pm$4.62 & & 7.83$\pm$7.02 & 3.11$\pm$2.59 & 2.76$\pm$2.70 \\
    {4C} & & - & 6.80$\pm$4.16 & 6.85$\pm$4.54 & & - & 3.92$\pm$3.29 & 3.76$\pm$3.09 \\
    \hline
    Mean & & - & 6.48$\pm$1.39 & \textbf{6.30}$\pm$1.26 & & - & 3.45$\pm$0.42 & \textbf{3.10}$\pm$0.57 \\
    \hline\hline
\end{tabular}
\end{adjustbox}
\end{table}

\subsubsection{Impact of Multi-View Aggregation}
We first investigate the impact of the proposed multi-view planning on plane prescription accuracy.
In theory, it is possible to define a target plane with only a subset of available source views, as the intersecting lines within two non-coinciding planes are sufficient to define any plane.
We examine such cases in our scenario and present the results in Table \ref{tab:multiview}.
The results show that despite being theoretically feasible, prescribing the target planes using only a subset of localizers leads to inferior performance in general, suggesting the importance of multi-view planning.
This is consistent with the clinical practice, where the operators consider multiple localizers to prescribe a standard view.
Notably, when using only the pLA (p2C and p4C) localizers to plan the 3C view, the performance collapses with a mean normal deviation of 41.88$^\circ$.
We speculate this is because the key landmark (\emph{i.e.}, the aortic valve) that defines the 3C view is only visible in the stack of pSA localizer.

\subsubsection{Impact of Gaussian Kernel Size}
Next, we study the impact of the Gaussian kernel size $\sigma$ in Equation (\ref{eq:heatmap}) by setting it to different values and comparing the validation performance.
Specifically, $\sigma$ is defined as a ratio/multiple of the slice thickness $t$ of the target view: $\sigma=\alpha\cdot t$, with $\alpha=$ 0.25, 0.5, 1, and 2.
The results are shown in Table \ref{tab:kernel_size}.
Based on a comprehensive evaluation of both metrics on all the four target planes, we choose $\sigma=0.5t$ for subsequent experiments.

\begin{table}[!t]
\caption{Impact of the Gaussian kernel size ($\sigma$) on plane prescription accuracy on the validation set.
$\sigma$ is defined as a ratio/multiple of the slice thickness $t$: $\sigma=\alpha\cdot t$ with $\alpha$ being a factor.
Data format: mean $\pm$ standard deviation.}\label{tab:kernel_size}
\centering
\setlength{\tabcolsep}{.7mm}
\begin{adjustbox}{width=\textwidth}
\begin{tabular}{rcccccccccc}
\hline\hline
\multicolumn{1}{c}{Target} & & \multicolumn{4}{c}{Normal deviation ($^{\circ}$)} & & \multicolumn{4}{c}{Point-to-plane distance (mm)} \\
\cline{3-6}\cline{8-11}
\multicolumn{1}{c}{plane} & $\sigma=$ & $0.25t$ & $0.5t$ & $1.0t$ & $2.0t$ & & $0.25t$ & $0.5t$ & $1.0t$ & $2.0t$ \\
\hline
{2C} & & 4.71$\pm$2.44 & 4.86$\pm$2.95 & 5.82$\pm$3.46 & 6.29$\pm$4.14 & & 2.66$\pm$2.02 & 2.78$\pm$1.92 & 3.47$\pm$2.51 & 4.88$\pm$3.23 \\
{3C} & & 7.14$\pm$4.18	& 7.18$\pm$4.62	& 7.42$\pm$4.90	& 7.97$\pm$5.29 & & 2.44$\pm$2.47	& 2.76$\pm$2.70	& 3.32$\pm$3.35	& 3.87$\pm$3.45 \\
{4C} & & 7.33$\pm$5.16	& 6.85$\pm$4.54	& 7.09$\pm$3.96	& 7.50$\pm$4.98 & & 3.76$\pm$3.03	& 3.76$\pm$3.09	& 3.63$\pm$2.83	& 4.59$\pm$3.56 \\
{SAX} & & 8.40$\pm$5.10 & 7.97$\pm$4.61	& 8.04$\pm$5.91	& 8.21$\pm$4.53 & & 3.11$\pm$2.37	& 3.07$\pm$2.45	& 3.77$\pm$2.87	& 4.38$\pm$2.57 \\
\hline
Mean & & 6.89$\pm$1.56 & \textbf{6.71}$\pm$1.32 & 7.09$\pm$0.94 & 7.49$\pm$0.85 & & \textbf{2.99}$\pm$0.58 & 3.09$\pm$0.47 & 3.55$\pm$0.20 & 4.43$\pm$0.43 \\
\hline\hline
\end{tabular}
\end{adjustbox}
\end{table}

\subsubsection{Evaluation on Test Set}
We now evaluate our proposed system on the held-out test set, and compare the performance with existing approaches \cite{alansary2018automatic,blansit2019deep,frick2011fully,lu2011automatic}.
The results are charted in Table \ref{tab:cmp}.
Above all, our method achieves the best mean performance averaged over the four standard CMR views in both evaluation metrics.
The second best method in terms of the mean normal deviation is \cite{blansit2019deep} (6.56$\pm$1.24$^\circ$ versus 5.98$\pm$0.79$^\circ$).
This method is similar to ours in that it also exploited the power of deep CNNs.
However, it required extensive annotations for training, whereas our method completely eliminates this demanding requirement and directly learns from data.
Further analysis of each specific target plane indicates that our method is also competent for each view: it yields the best point-to-plane distances for all the four standard views, the lowest normal deviations for the 2C and 3C views, and the second lowest normal deviations for the 4C and SAX views.
Lastly, our method can also successfully prescribe the pseudo-view localizers from the axial localizer, with a mean performance comparable to that on prescribing the standard CMR views.
Prescribing these localizers is the first step from body-oriented imaging planes to the cardiac anatomy oriented imaging planes of CMR.
So far as the authors are aware of, it has not been previously demonstrated for any clinic-compatible system.
These results demonstrate the competence and effectiveness of our proposed system in automatic CMR view planning.
Examples of the automatic prescription results are shown in Fig.~\ref{fig:results} (more examples in supplement Fig. \ref*{figS:results}).

\begin{table}[t]
  \caption{Test set evaluation results and comparison to previous works.
  Bold and italic fonts indicate best and second best results, respectively.
  NA: not applicable.
  -: not reported.
  *: significance at 0.05 level for pairwise comparison to the proposed method (Bonferroni correction applied where appropriate);
  since different works used different data, independent samples \emph{t}-test is employed.
  Data format: mean $\pm$ standard deviation.}\label{tab:cmp}
  \centering
  \setlength{\tabcolsep}{1.78mm}
  \begin{adjustbox}{width=\textwidth}
  \begin{threeparttable}
  \begin{tabular}{r|ccccc|c}
    \hline\hline
    Methods & 2C & 3C & 4C & SAX & Mean & Localizers\textsuperscript{a} \\
    \Xhline{1pt}
    \multicolumn{7}{c}{\emph{Normal deviation} ($^\circ$)} \\
    \hline
    Lu \emph{et al.} \cite{lu2011automatic} & 18.9$\pm$21.0* & 12.3$\pm$11.0* & 17.6$\pm$19.2* & 8.6$\pm$9.7 & 14.35$\pm$4.78 & NA \\
    Alansary \emph{et al.} \cite{alansary2018automatic} & - & - & 8.72$\pm$7.44 & - & - & NA \\
    Frick \emph{et al.} \cite{frick2011fully} & 7.1$\pm$3.6* & 9.1$\pm$6.3 & 7.7$\pm$6.1 & 6.7$\pm$3.6 & 7.65$\pm$1.05 & NA \\
    Blansit \emph{et al.} \cite{blansit2019deep} & \emph{8.00}$\pm$6.03* & \emph{7.19}$\pm$4.97 & \textbf{5.49}$\pm$5.06 & \textbf{5.56}$\pm$4.60 & \emph{6.56}$\pm$1.24 & - \\
    Proposed & \textbf{4.97}$\pm$4.00 & \textbf{6.84}$\pm$4.16 & \textit{5.84}$\pm$3.19 & \emph{6.28}$\pm$3.48 & \textbf{5.98}$\pm$0.79 & 5.88$\pm$0.53 \\
    \Xhline{1pt}
    \multicolumn{7}{c}{\emph{Point-to-plane distance} (mm)} \\
    \hline
    Lu \emph{et al.} \cite{lu2011automatic} & 6.6$\pm$8.8* & 4.6$\pm$7.7 & 5.7$\pm$8.5 & 13.3$\pm$16.7* & 7.55$\pm$3.92 & NA \\
    Alansary \emph{et al.} \cite{alansary2018automatic} & - & - & \emph{5.07}$\pm$3.33* & - & - & NA \\
    Proposed & \textbf{2.68}$\pm$2.34 & \textbf{3.44}$\pm$2.37 & \textbf{3.61}$\pm$2.63 & \textbf{4.18}$\pm$3.15 & \textbf{3.48}$\pm$0.62 & 4.18$\pm$0.50 \\
    \hline\hline
  \end{tabular}
  \begin{tablenotes}
    \item[a] Mean across the localizers; separate results for each localizer are presented in supplement Table \ref*{tabS:loc_results}.
  \end{tablenotes}
  \end{threeparttable}
  \end{adjustbox}
\end{table}

\begin{figure}[t]
\centering
\includegraphics[height=.214\textwidth,trim=0 0 0 149,clip]{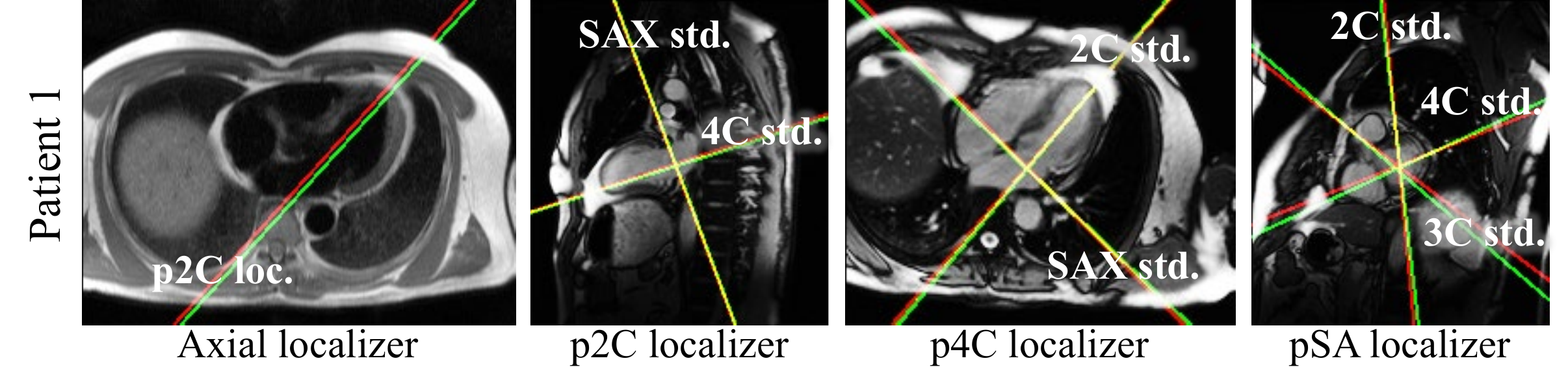}
\caption{Example CMR view plane prescription results by the proposed system.
Green: ground truth; red: automatic prescription; yellow: overlap.
Due to the closeness (sometimes coincidence) of the standard (std.) 3C view with std. 2C and 4C views in the p4C and p2C localizers (loc.), the std. 3C view is visualized only in the pSA loc.}\label{fig:results}
\end{figure}

\section{Conclusion}
In this work, we proposed a CNN-based, clinic-compatible system for automatic CMR view planning.
Our system was distinctive from a closely related work in that it eliminated the burdensome need for manual annotations, by mining the spatial relationship between CMR views.
Also, the importance of the proposed multi-view aggregation---an analogue to the behaviour of human prescribers---was empirically validated.
Experimental results showed that our system was superior to existing approaches including both conventional atlas-based and newer CNN-based ones, in prescription accuracy of four standard CMR views.
In addition, we demonstrated accurate prescription of the pseudo-view localizers from the axial localizer and filled the gap in existing literature.
Lastly, based on the encouraging results, we believe our work has opened up a new direction for automatic view planning of anatomy-oriented medical imaging beyond CMR.

\subsubsection{Acknowledgments.} This work was supported by the Key-Area Research and Development Program of Guangdong Province, China (No. 2018B010111001), and Scientific and Technical Innovation 2030 - ``New Generation Artificial Intelligence'' Project (No. 2020AAA0104100).
%
%
%
\bibliographystyle{splncs04}
\bibliography{mybibliography}
%
%
%
%
%
%
\newpage
\begin{center}
\textbf{\large Supplementary Material: Training Automatic View Planner for Cardiac MR Imaging via Self-Supervision by Spatial Relationship between Views}
\end{center}
\setcounter{equation}{0}
\setcounter{figure}{0}
\setcounter{table}{0}
\setcounter{page}{1}
\makeatletter
\renewcommand{\theequation}{S\arabic{equation}}
\renewcommand{\thefigure}{S\arabic{figure}}
\renewcommand{\thetable}{S\arabic{table}}
\renewcommand{\thepage}{S\arabic{page}}
%
\begin{figure}[h]
\centering
\includegraphics[width=.888\textwidth,trim=0 0 0 0,clip]{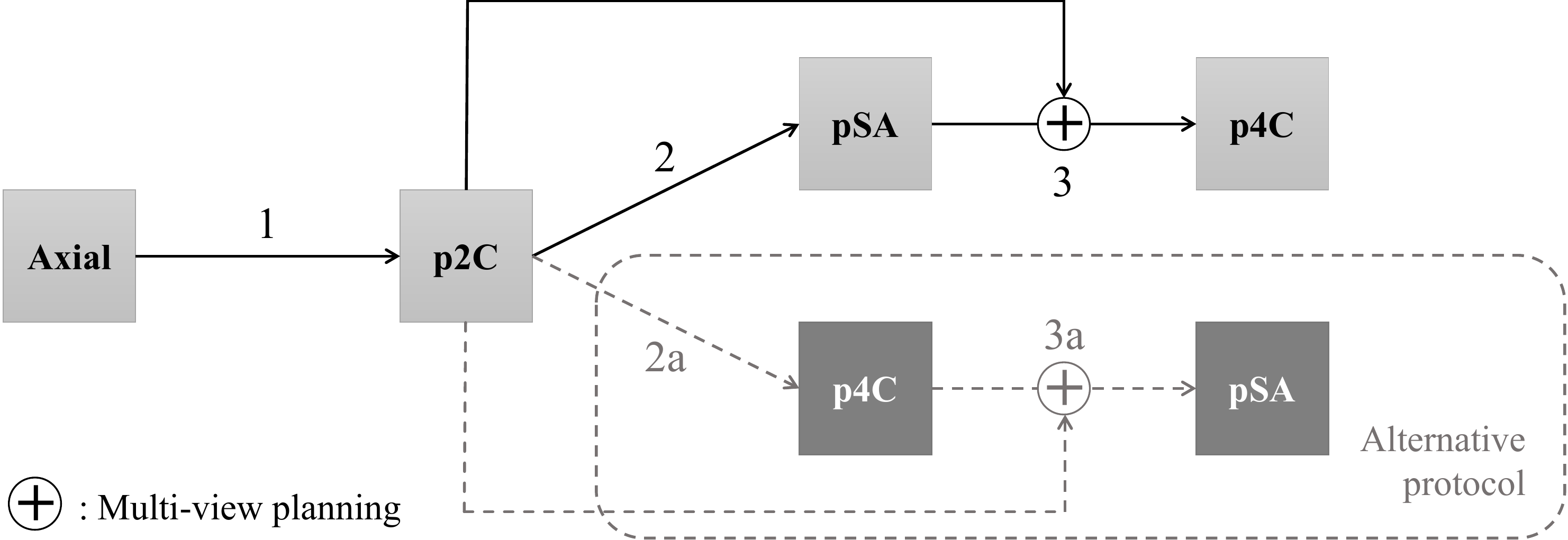}
\caption{Our protocol for planning the set of pseudo-view localizers, starting from the axial localizer.
Step 1: prescribe the p2C localizer orthogonal to the axial localizer.
Step 2: prescribe the pSA localizer orthogonal to the p2C localizer.
Step 3: prescribe the p4C localizer via multi-view planning by referring to both the p2C and pSA localizers; usually the resulting p4C localizer is oblique to both the p2C and pSA localizers.
An alternative protocol (used in 40\% of the 181 CMR exams) is illustrated in Steps 2a and 3a, where the order of prescribing the p4C and pSA localizers is switched.} \label{figS:loc}
\end{figure}

\begin{table}[h]
  \caption{Plane prescription dependency relationships between the CMR views in our protocol.
  The relationships are also illustrated in Fig. \ref*{figS:loc} (for the localizers) and Fig.~\ref{fig:framework} (for the standard views).}\label{tabS:view_dependency}
  \centering
  \setlength{\tabcolsep}{.7mm}
  \begin{adjustbox}{width=\textwidth}
  \begin{threeparttable}
  \begin{tabular}{c|c|c|c|c|c|c|c}
    \hline\hline
    Target plane & p2C & pSA\textsuperscript{a} & p4C\textsuperscript{a} & 2C & 3C & 4C & SAX \\
    \hline
    Source view & Axial & p2C (\& p4C) & p2C \& pSA (p2C) & p4C \& pSA & p2C, p4C \& pSA & p2C \& pSA & p2C \& p4C\\
    \hline\hline
  \end{tabular}
  \begin{tablenotes}
    \item[a] The source views in parentheses correspond to the alternative protocol (Fig. \ref*{figS:loc} 2a and 3a).
  \end{tablenotes}
  \end{threeparttable}
  \end{adjustbox}
\end{table}

\begin{table}[!hb]
  \caption{Test set evaluation results for the pseudo-view localizers.
  Note that for prescription of the p4C and pSA localizers we reuse the CNN models trained for the standard 4C and SAX views for heatmap prediction.
  Also, when planning a target plane orthogonal to its only source view (\emph{e.g.}, Steps 2 and 2a in Fig. \ref*{figS:loc}), the problem of searching for the optimal target plane degenerates into a simpler one of searching for an optimal line in the source view.
  Data format: mean $\pm$ standard deviation.}\label{tabS:loc_results}
  \centering
  \setlength{\tabcolsep}{2.5mm}
  \begin{adjustbox}{width=.9\textwidth}
  \begin{tabular}{r|ccc|c}
    \hline\hline
    Localizer & p2C & p4C & pSA & Mean\\
    \hline
    Normal deviation ($^\circ$) & 6.31$\pm$3.97 & 5.29$\pm$3.87 & 6.04$\pm$4.06 & 5.88$\pm$0.53 \\
    Point-to-plane distance (mm) & 4.11$\pm$3.42 & 4.72$\pm$3.08 & 3.72$\pm$1.37 & 4.18$\pm$0.50 \\
    \hline\hline
  \end{tabular}
  \end{adjustbox}
\end{table}


\begin{table}[t]
  \caption{Sequence parameters used for the CMR image acquisition.
  All images were acquired with the 1.5T Siemens Symphony model.
  Abbreviations: HASTE: half-Fourier acquisition single-shot turbo spin-echo;
  TRUFI: true fast imaging with steady-state free precession;
  FISP: fast imaging with steady-state free precession.}\label{tabS:protocol}
  \centering
  \setlength{\tabcolsep}{.65mm}
  \begin{adjustbox}{width=\textwidth}
  \begin{threeparttable}
  \begin{tabular}{rcccccccc}
    \hline\hline
    \multicolumn{1}{c}{View} & Axial & p2C & p4C & pSA & 2C & 3C & 4C & SAX \\
    \hline
    \multicolumn{1}{c}{\multirow{ 2}{*}{Sequence}} & HASTE & \multirow{ 2}{*}{TRUFI} & \multirow{ 2}{*}{TRUFI} & \multirow{ 2}{*}{TRUFI} & \multicolumn{2}{c}{\multirow{ 2}{*}{TrueFISP cine retro}} & \multicolumn{2}{c}{\multirow{ 2}{*}{TrueFISP cine retro}} \\
    & dark blood \\
    \hline
    \multicolumn{1}{l}{Flip angle ($^\circ$)} \\
    \cline{1-1}
    Mode/Mean\textsuperscript{a} & 160 (100\%) & 65 (100\%) & 65 (100\%) & 65 (99\%) & 73 (83\%) & 73 (86\%) & 74 (79\%) & 73 (84\%) \\
    Range & - & - & - & 64, 65 & 49--79 & 49--79 & 49--80 & 53--79 \\
    \hline
    \multicolumn{1}{l}{Rows (pixel)} \\
    \cline{1-1}
    Mode/Mean\textsuperscript{a} & 192 (88\%) & 192 (100\%) & 160 (85\%) & 192 (99\%) & 192 (100\%) & 170 & 155 & 192 (98\%) \\
    Range & 176--224 & - & 128--192 & 160, 176, 192 & - & 132--192 & 132--192 & 156, 162, 192 \\
    \hline
    \multicolumn{1}{l}{Columns (pixel)} \\
    \cline{1-1}
    Mode/Mean\textsuperscript{a} & 256 (100\%) & 176 (73\%) & 192 (97\%) & 176 (82\%) & 156 & 173 & 192 (98\%) & 161 \\
    Range & - & 144--192 & 160, 176, 192 & 144--192 & 144--180 & 132--192 & 144--192 & 144--192 \\
    \hline
    \multicolumn{1}{l}{Pixel spacing\textsuperscript{b} (mm)} \\
    \cline{1-1}
    Mode/Mean\textsuperscript{a} & 1.34 & 1.98 (66\%) & 1.77 (90\%) & 1.88 (95\%) & 1.80 & 1.67 (72\%) & 1.72 & 1.85 \\
    Range & 1.17--1.56 & 1.67--2.29 & 1.56--2.08 & 1.77--1.98 & 1.51--2.08 & 1.46--2.08 & 1.56--1.93 & 1.56--2.08 \\
    \hline
    \multicolumn{1}{l}{Slice thickness (mm)} \\
    \cline{1-1}
    Mode & 6 (100\%) & 6 (100\%) & 6 (100\%) & 6 (88\%) & 7 (96\%) & 7 (96\%) & 7 (96\%) & 7 (96\%) \\
    Range & - & - & - & 6, 6.5, 7 & 7, 8 & 6, 7, 8 & 7, 8 & 7, 8 \\
    \hline
    \multicolumn{1}{l}{Repetition time (ms)} \\
    \cline{1-1}
    Mode/Mean\textsuperscript{a} & 731 & 437 & 437 (93\%) & 426 (96\%) & 42.22 & 43.16 & 42.66 & 41.94 \\
    Range & 470--1000 & 400--798 & 371--840 & 423--857 & 40.18--63.8 & 40.46--51.04 & 41.02--63.2 & 39.9--45.08 \\
    \hline
    \multicolumn{1}{l}{Echo time (ms)} \\
    \cline{1-1}
    Mode/Mean\textsuperscript{a} & 28 (99\%) & 1.24 (89\%) & 1.27 (94\%) & 1.26 (95\%) & 1.28 & 1.31 & 1.29 & 1.27 \\
    Range & 28, 38 & 1.19--1.30 & 1.21--1.34 & 1.23--1.29 & 1.21--1.36 & 1.21--1.39 & 1.24--1.34 & 1.21--1.37 \\
    \hline
    \multicolumn{1}{l}{Number of slices} \\
    \cline{1-1}
    Mode/Mean\textsuperscript{a} & 30 (89\%) & 1 (100\%) & 1 (100\%) & 8 (83\%) & 1 (100\%) & 1 (100\%) & 1 (100\%) & 10.7 \\
    Range & 30--37 & - & - & 6--10 & - & - & - & 3--14 \\
    \hline
    \multicolumn{1}{l}{Slice interval\textsuperscript{c} (mm)} \\
    \cline{1-1}
    Mode/Mean\textsuperscript{a} & 6 (99\%) & - & - & 12 (85\%) & - & - & - & 10 (99\%) \\
    Range & 6, 6.6 & - & - & 10.8--15 & - & - & - & 5, 9.95, 10 \\
    \hline\hline
  \end{tabular}
  \begin{tablenotes}
    \item[a] The mode is shown if it occupies the vast majority of the dataset (with the ratio in parentheses), otherwise the mean is shown.
    \item[b] Pixels are isotropic.
    \item[c] Measured from center-to-center of each slice along the common normal to the stack of slices.
  \end{tablenotes}
  \end{threeparttable}
  \end{adjustbox}
\end{table}

\begin{figure}[b]
\centering
\includegraphics[height=.214\textwidth,trim=0 141 0 1,clip]{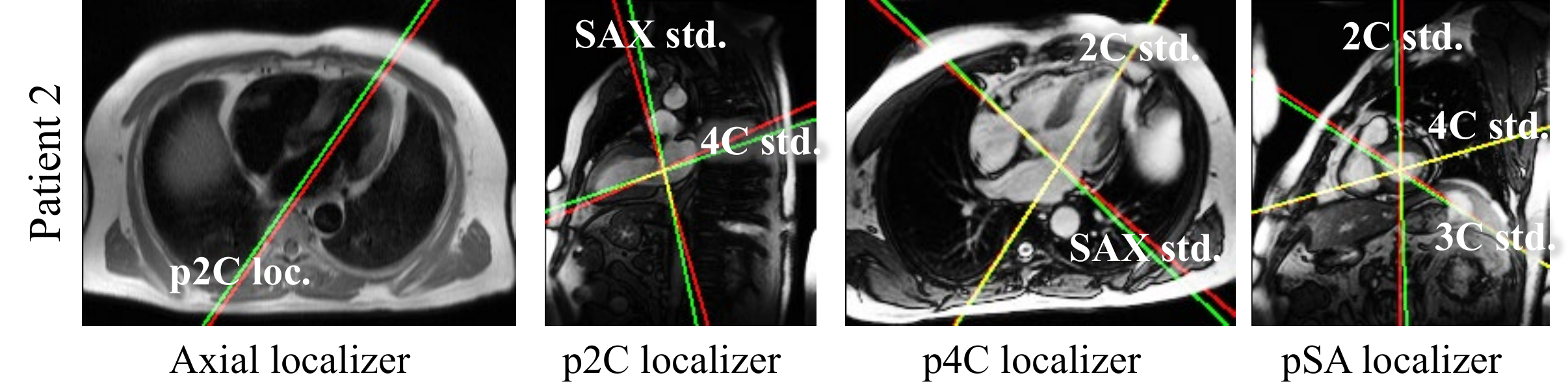}
\caption{Example CMR view plane prescription results by the proposed system for a different patient from Fig. \ref{fig:results}.
Green: ground truth; red: automatic prescription; yellow: overlap.
Due to the closeness (sometimes coincidence) of the standard (std.) 3C view with std. 2C and 4C views in the p4C and p2C localizers (loc.), the std. 3C view is visualized only in the pSA loc.}\label{figS:results}
\end{figure}
\end{document}